\\

Title: Using of the natural radioactive elements for determining Ge-detector efficiencies

Autors: E. G. Tertyshnik, I. E. Epifanova

A method is proposed to use of the mixture of Uran oxide and non-active matrix (e.g., NaCl) and also potassium and lanthanum for determining Ge-detector efficiencies. The preparations containing of known amouts of the U or K, or La were measured by means of the Ge-detector, which a efficiency curve has been obtained through the reference solutions of $^{241}$Am, $^{109}$Cd, $^{57}$Co, $^{139}$Ce, $^{137}$Cs, $^{60}$Co. Results the measurements were compared the activities of the preparations calculated from mass of $^{235}$U, $^{238}$U, $^{138}$La, $^{40}$K in the samples, its natural abundance, half lives and intensities of gamma lines. Discrepancy of the activities in the energy range between 163 and 1461 keV does not exeed 7 %.

For correct comparison of the activities the coefficients $\omega$ were calculated, which took into consideration a varied sorption of gamma-rays in water and in mixture of the Uran oxide and matrix.

--------------------------------------------------------

Заголовок: Применение естественных радиоактивных элементов для калибровки гамма-спектрометров по эффективности регистрации

Авторы: Э. Г. Тертышник, И. Э. Епифанова

Предложен метод использования смеси оксида урана с неактивной матрицей (например, NaCl), а также калия и лантана для определения эффективности регистрации Ge- детектора. Препараты, содержащие известное количество U или K, или La измерялись с помощью Ge- детектора, кривая эффективности которого была получена путём измерения Образцовых радиоактивных растворов (ОРР) $^{241}$Am, $^{109}$Cd, $^{57}$Co, $^{139}$Ce, $^{137}$Cs, $^{60}$Co. Результаты измерений сравнивали с активностями препаратов, рассчитанными по массе $^{235}$U, $^{238}$U, $^{138}$La, $^{40}$K в препаратах, содержанию в природной смеси, периодам полураспада и интенсивностью соответствующих гамма-линий. Расхождение результатов в диапазоне энергий гамма-квантов 163 – 1461 кэВ не превышало 7 %.

Для корректного сравнения активностей коэффициенты $\omega$ были рассчитаны, чтобы учесть различное поглощение гамма-квантов в воде (ОРР) и смеси оксида урана и хлористого натрия.

\\

# Применение естественных радиоактивных элементов для калибровки гамма-спектрометров по эффективности регистрации

Тертышник Э.Г.( ФГБУ НПО "Тайфун", г. Обнинск, Калужской обл. ),

Епифанова И.Э. (ГНУ ВНИИСХРАЭ Россельхозакадемии, г. Обнинск)

Контроль радиоактивного загрязнения окружающей среды предполагает широкое использование гамма-спектрометров, калибровка которых по эффективности регистрации – необходимое условие корректного определения гамма-излучателей в природных средах. Для калибровки применяют поставляемые метрологическими организациями Образцовые Радиоактивные растворы (ОРР), содержащие один или несколько радионуклидов. Аккуратное проведение операций разбавления, фасовки и взвешивания радиоактивных растворов обеспечивает калибровку гамма-спектрометра с относительной погрешностью 3 – 5% .Однако применение ОРР для калибровки имеет следующие недостатки:

Калибровку, как правило, невозможно повторить через некоторое время, т.к. изготовитель не гарантирует продолжительное сохранение аттестованных параметров после вскрытия ампулы с ОРР ;

стоимость ОРР высока и повышается пропорционально числу аттестованных радионуклидов. Радионуклиды, испускающие гамма-кванты разных энергий, в частности $^{152}$Eu, $^{133}$Ba, не пригодны для моделирования «близких» геометрий, когда измеряемый препарат размещается непосредственно на детекторе, вследствие эффекта гамма-гамма совпадений;

при работе с жидкими препаратами существует вероятность случайного радиоактивного загрязнения детектора, которое практически невозможно удалить.

Чтобы устранить эти недостатки, было предложено в хорошо оснащённой лаборатории готовить калибровочные препараты путем введения ОРР в различные неорганические (песок, почву, цеолиты) и органические (манную крупу, сахар и др.) материалы [1]. После соответствующего разбавления ОРР вносили или непосредственно в матрицу препарата или наносили на небольшое количество активированного угля, затем этот уголь тщательно перемешивали с матрицей. Полученным материалом заполняли стандартные контейнеры и после их герметизации готовые рабочие эталоны направляли в лаборатории, выполняющие рутинные анализы.

В настоящей работе предлагается рабочие эталоны для калибровки гамма-спектрометров по эффективности готовить путем смешивания оксидов урана с неактивной матрицей. Принципиальная возможность использования такого метода калибровки не вызывает сомнений, но его практическая реализация зависит от того, насколько имеющиеся данные о периодах полураспада и значениях квантовых

Активность радионуклида (Бк) в приготовленной смеси с неактивной матрицей однозначно определяется его массой и находится по формуле:

$$Q = \lambda N = \lambda m N_A / A , \qquad (1)$$

где

N - число атомов данного радионуклида в препарате;

$\lambda$ - постоянная распада радионуклида, $с^{-1}$;

m - масса радионуклида в препарате, г;

A - массовое число;

$N_A$ - число Авогадро $6{,}022 \cdot 10^{23}$ моль$^{-1}$

В табл.1 приведена удельная активность природных элементов: лантана, калия и урана, рассчитанная по формуле (1). Периоды полураспада взяты из [2], а массовая доля радионуклидов в природной смеси изотопов данного элемента из [3].

Т а б л и ц а 1. Удельная активность некоторых элементов и их гамма-излучающие изотопы

| Изотоп | Массовая доля изотопа в природной смеси, % [3] | Период полураспада, лет [2] | Постоянная распада, $с^{-1}$ | Удельная активность, Бк/г | Удельная активность изотопа в природной смеси q, Бк/г |
|---|---|---|---|---|---|
| $^{138}$La | 0,089 | $1{,}35 \cdot 10^{11}$ | $1{,}627 \cdot 10^{-19}$ | $7{,}10 \cdot 10^{2}$ | 0,632 |
| $^{40}$K | 0,0119 | $1{,}277 \cdot 10^{9}$ | $1{,}720 \cdot 10^{-17}$ | $2{,}59 \cdot 10^{5}$ | 30,8 |
| $^{235}$U | 0,7196 | $7{,}038 \cdot 10^{8}$ | $3{,}121 \cdot 10^{-17}$ | $8{,}00 \cdot 10^{4}$ | 576 |
| $^{238}$U | 99,274 | $4{,}468 \cdot 10^{9}$ | $4{,}918 \cdot 10^{-18}$ | $1{,}244 \cdot 10^{4}$ | $1{,}235 \cdot 10^{4}$ |

Для проверки возможности применения предлагаемого метода калибровки были приготовлены препараты на основе лантана, калия и природного урана. Для взвешивания использовали лабораторные весы типа АДВ-200, объем определяли с помощью мерного цилиндра (250 мл, 2-го класса). Затем проведен анализ препаратов на гамма-спектрометре, предварительно калиброванном с помощью

ОРР. Сравнение полученных экспериментальных результатов с расчетными (табл.1) позволяет судить о возможностях метода.

Препарат, содержащий лантан, был изготовлен из предварительно прокаленной и измельченной азотнокислой соли лантана $La(NO_3)_3$, и помещен в стандартный цилиндрический контейнер из полиэтилена. Внутренний диаметр контейнера – 72 мм, толщина дна – 1 мм. Препарат массой 155 г (объем 110 мл) содержал 66,26 г лантана.

Для приготовления препарата, содержащего калий, использовали химически чистый сухой хлористый калий (KCl), масса которого составила 110,7 г (масса калия 58,0 г), объем, занимаемый солью в стандартном измерительном контейнере, - 110 мл.

Препарат, содержащий уран, приготовили смешиванием хлористого натрия (NaCl 141,1 г) и 6,01 г диоксида природного урана. В смеси, помещенной в стандартный контейнер объемом 110 мл, содержание урана составило 5,298 г.

Препараты измеряли на γ-спектрометре с детектором из сверхчистого германия LO-AX-60495 фирмы "EG&G ORTEC" (США), диаметр германиевого кристалла 60 мм, толщина около 29 мм и амплитудного анализатора SBS - 50 фирмы "Зеленая Звезда", г.Москва. Калибровка гамма-спектрометра по эффективности для разных геометрий измерения была проведена с помощью ОРР, в растворе присутствовали аттестованные по удельной активности $^{241}Am, ^{109}Cd, ^{57}Co, ^{139}Ce, ^{113}Sn, ^{137}Cs, ^{88}Y, ^{60}Co$; согласно паспорту ОРР№ 601/99/20608 погрешность аттестации не превышала 2% (Р=0,95). Полученная при калибровке зависимость эффективности от энергии гамма-квантов для стандартного контейнера объемом 110мл, размещенного на детекторе, (при энергии свыше 160кэВ) хорошо аппроксимируется формулой

$$\varepsilon_0(E) = 14{,}953 E^{-1{,}074} + 3{,}4 \cdot 10^{-4}, \qquad (2)$$

где Е - энергия γ-квантов, кэВ.

Для такого же препарата, удаленного от детектора на 100 мм, аппроксимирующее выражение имеет вид

$$\varepsilon_0(E) = 1{,}817 E^{-1{,}105} + 4{,}5 \cdot 10^{-5}$$

Стабильность эффективности регистрации во времени контролировали с помощью рабочего эталона 2-го разряда (объемный сыпучий гранулированный источник гамма-излучения № 57/01), который содержал $^{133}Ba, ^{137}Cs, ^{60}Co$. Суммарная погрешность удельной активности согласно паспорту не превышала 7% (Р = 0,95).

Активность радионуклида (Бк) в препаратах рассчитывали по формуле

$$Q = n / \eta_i \, \varepsilon_{0i} \, \omega_i , \qquad (3)$$

где n - скорость счета импульсов в соответствующем пике полного поглощения, $с^{-1}$;

$\eta_i$, $\varepsilon_{0i}$ - квантовый выход и эффективность регистрации i-й гамма-линии соответственно (в нашем случае эффективность регистрации получена при калибровке спектрометра с помощью ОРР, т.е. матрицей образцового источника является вода); $\omega_i$ - коэффициент, учитывающий различное ослабление гамма-квантов i-й энергии в материале (матрице) препарата и в материале образцового источника, с помощью которого калибровали гамма-спектрометр по эффективности регистрации . Этот коэффициент можно рассчитать по формуле [4]

$$\omega = \mu_0\rho_0[1 - \exp(1 - \mu\rho h)] / [\mu\rho(1 - \exp(1 - \mu_0\rho_0 h))] , \qquad (4)$$

где $\mu_0(E)$, $\mu(E)$ - массовый коэффициент ослабления излучения для материала образцового источника и пробы соответственно, $см^2/г$ ;

$\rho_0$, $\rho$ - плотность материала эталонного источника и насыпная плотность препарата соответственно, $г/см^3$ ;

h - толщина (высота) слоя препарата в измерительном контейнере, см .

В нашем случае калибровки по ОРР материал образцового источника - вода ($\rho_0$=1 $г/см^3$ , h=2,7 см) .

В табл.2 приведены результаты расчета поправочного коэффициента $\omega$ по формуле (4) для гамма-квантов, излучаемых ураном. Здесь же представлены значения $\mu(E)$ для натрия, хлора, урана и значения $\mu_0(E)$, полученные путем интерполяции справочных данных [5]. Массовый коэффициент ослабления излучения для смеси хлористого натрия и $UO_2$ определяли по формуле

$$\mu_m = P_1\mu_1 + P_2\mu_2 + P_3\mu_3 ,$$

где $P_1$, $P_2$, $P_3$ — содержание в смеси натрия, хлора и урана соответственно, ($P_1+P_2+P_3 =1$);

$\mu_1$, $\mu_2$, $\mu_3$ — массовый коэффициент ослабления для натрия, хлора и урана, соответственно.

Поскольку значения массового коэффициента ослабления излучения для кислорода и хлористого натрия отличаются незначительно, доля кислорода (в диоксиде урана) при расчётах была включена в долю хлористого натрия.

Препараты, содержащие лантан и калий, во время измерений устанавливались на торцевую поверхность детектора, закрытую защитным колпаком из оргстекла (толщина торцевой стенки колпака 1 мм), который предохранял бериллиевое окно детектора от повреждений. Эффективность регистрации для соответствующих гамма-линий рассчитывали по формуле (2). Препарат, содержащий уран, устанавливали на расстоянии 100 мм от торцевой поверхности детектора (с помощью фиксирующей втулки), чтобы исключить возможное влияние эффекта гамма-гамма совпадений на результаты измерений. В случае определения активности радионуклидов по нескольким аналитическим линиям усреднение активности проводили с учетом веса

$$Q = \Sigma Q_i W_i / \Sigma W_i ,$$

где - $Q_i$ - активность, рассчитанная по $i$-й линии;

$W_i$ весовой множитель, в качестве которого использовалась скорость счета импульсов в пике полного поглощения, соответствующем $i$-й энергии гамма-линии.

Т а б л и ц а 2. Поправочный коэффициент $\omega$ для препарата, содержащего диоксид урана (6,01г) и хлористый натрий (141,1г)

| Энергия γ-квантов, кэВ | Массовый коэффициент ослабления излучения, см²/г (Содержание элемента в смеси) [5] | | | | | Поправочный коэффициент, $\omega$ |
|---|---|---|---|---|---|---|
| | Натрий (0,379) | Хлор (0,584) | Уран (0,036) | Смесь NaCl, UO$_2$ | Вода | |
| 63,3 | 0,19 | 0,34 | 5,49 | 0,47 | 0,187 | 0,62 |
| 92,6 | 0,16 | 0,20 | 2,02 | 0,25 | 0,17 | 0,83 |
| 143,8 | 0,13 | 0,14 | 2,7 | 0,23 | 0,15 | 0,83 |
| 163,4 | 0,12 | 0,13 | 2,0 | 0,19 | 0,14 | 0,87 |
| 185,7 | 0,12 | 0,12 | 1,48 | 0,17 | 0,14 | 0,90 |
| 205.3 | 0,11 | 0,12 | 1,17 | 0,15 | 0,13 | 0,92 |
| 766 | 0,071 | 0,070 | 0,094 | 0,071 | 0,081 | 0,98 |
| 1001 | 0,061 | 0,061 | 0,073 | 0,061 | 0,072 | 0,99 |

Результаты измерения препаратов и использованные при расчете активности по формуле (3) значения квантовых выходов и эффективности регистрации приведены в табл.3 .

Результирующая относительная погрешность определения активности, обусловленная погрешностью калибровки гамма-спектрометра по эффективности, неточностью воспроизведения геометрии измерения и погрешностью определения площади пика полного поглощения, по нашим оценкам не превысила 5 ─ 7% (при доверительной вероятности 0,95). В случае измерения препаратов, содержащих уран, может вноситься дополнительная погрешность за счет несовершенной гомогенизации при изготовлении препаратов.

Из табл.3 видно, что имеет место хорошее совпадение результатов измерения и расчета удельной активности (табл.1) для $^{40}$K и $^{138}$La (по гамма-линии 788,4 кэВ). При расчете удельной активности лантана по линии 1436 кэВ она оказывается на 7% ниже расчетной.

Т а б л и ц а 3. Результаты гамма-анализа препаратов на основе лантана, калия и урана

| Радио-нуклид | E, кэВ | n, имп./с | η | $\varepsilon_0$ | ω | Q, Бк | s=Q/m, Бк/г | s/q |
|---|---|---|---|---|---|---|---|---|
| $^{138}$La | 788,4 | 0,162 | 0,329 [2] | $1,19 \cdot 10^{-2}$ | 0,98 | 42,2 | 0,64 | 1,01 |
| $^{138}$La | 1436 | 0,165 | 0,671 [2] | $6,41 \cdot 10^{-3}$ | 0,99 | 38,7 | 0,58 | 0,93 |
| $^{40}$K | 1461 | 1,213 | 0,107 [2] | $6,30 \cdot 10^{-3}$ | 1,01 | $1,78 \cdot 10^3$ | 30,7 | 1,0 |
| | | | | | | | | |
| $^{235}$U | 143,8 | 1,78 | 0,1096 [6,7] | $7,40 \cdot 10^{-3}$ | 0,83 | $2,64 \cdot 10^3$ | 499 | |
| $^{235}$U | 163,3 | 0,822 | 0,0508 [6,7] | $6,50 \cdot 10^{-3}$ | 0,87 | $2,86 \cdot 10^3$ | 540 | |
| $^{235}$U | 185,7 | 9,05 | 0,572 [6,7] | $5,70 \cdot 10^{-3}$ | 0,90 | $3,08 \cdot 10^3$ | 582 | |
| $^{235}$U | 205.3 | 0.712 | 0,0501 [6,7] | $5,10 \cdot 10^{-3}$ | 0,92 | $3,03 \cdot 10^3$ | 572 | |
| $^{235}$U | Средневзвешенное значение | | | | | $3,00 \cdot 10^3$ | 566 | 0,98 |
| $^{234}$Th | 63,3 | 13,09 | 0,0484 [6,7] | $9,3 \cdot 10^{-3}$ | 0,62 | $4,69 \cdot 10^4$ | $8,86 \cdot 10^3$ | 0,72 |
| $^{234}$Th | 92,4 | 21,94 | 0,0558 [6,7] | $9,8 \cdot 10^{-3}$ | 0,83 | $4,83 \cdot 10^4$ | $9,12 \cdot 10^3$ | 0,74 |
| $^{234m}$Pa | 766 | 0,222 | ,00294[6,7] | $1,23 \cdot 10^{-3}$ | 0,98 | $6,26 \cdot 10^4$ | $1,18 \cdot 10^4$ | |
| $^{234m}$Pa | 1001 | 0,44 | ,00837[6,7] | $0,92 \cdot 10^{-3}$ | 0.99 | $5,77 \cdot 10^4$ | $1,09 \cdot 10^4$ | |
| $^{234m}$Pa | Средневзвешенное по линиям 766 и 1001кэВ | | | | | $5,92 \cdot 10^4$ | $1,12 \cdot 10^4$ | 0,91 |

Поскольку относительный выход гамма-линий определяется с хорошей точностью, можно предположить, что период полураспада $^{138}$La для процесса K-захвата в результате которого испускаются гамма-кванты 1436 кэВ, немного отличен от периода полураспада этого радионуклида путём β-распада, сопровождающимся испусканием квантов с энергией 788,4 кэВ.

Измерение удельной активности $^{234}$Th по линиям 63,3 и 92,6 кэВ дали значения на 26% ниже удельной активности $^{238}$U, рассчитаной по массе урана в препарате. Эти активности должна быть равны, так как $^{238}$U находится в состоянии радиоактивного равновесия с дочерним $^{234}$Th и $^{234m}$Pa. Заметим, что если использовать данные по квантовым выходам, приводимым в [8] η (63,3) = 0,037 и η (92,4 + 92,8) = 0,0423, то результаты измерений и расчёта совпадают с точностью 6 - 3 %.

Хотя при измерении активности $^{234m}$Pa по линиям 766 и 1001 кэВ результаты оказались примерно на 10% меньше расчетных, целесообразно использовать эти линии для калибровки, принимая во внимание почти 100%-ное содержание $^{238}$U в урановых препаратах, в то время как содержание $^{235}$U может существенно меняться, если используются обеденные (или обогащенные) по $^{235}$U соединения урана. В природном уране отношение содержания $^{235}$U к $^{238}$U составляет 1:137 (табл.1), в случае использования соединений обеденного (обогащенного) урана это отношение необходимо измерить путем масс-спектрометрии.

Для $^{235}$U измеренная и рассчитанная активность согласуются при определении по линиям 163,4 , 185,7 и 205,3 кэВ. Определение активности $^{235}$U по гамма-линии 143,8 кэВ приводит к занижению активности на 14%.

Таким образом экспериментально показано, что калибровку гамма-детекторов по эффективности регистрации можно проводиться с помощью препаратов калия, лантана и урана в диапазоне энергий гамма-квантов от 163 до 1461кэВ, если рассчитывать активность $^{40}$K, $^{138}$La, $^{235}$U и $^{238}$U по массе элементов в препаратах, применяя приведенные константы (период полураспада, содержание радионуклида в природной смеси данного радиоэлемента, квантовый выход используемых гамма-линий ). В работе [9] для калибровки германиевых детекторов кроме калия и лантана предлагается использовать лютеций, который содержит долгоживущий гамма-излучатель $^{176}$Lu (энергии гамма-линий 88, 202 и 307 кэВ).

В качестве примера в табл.4 приведены результаты калибровки гамма-спектрометра с детектором марки GEM-30185 фирмы EG&G ORTEC с помощью описанных препаратов калия, лантана и урана. Дополнительно использовали препарат, приготовленный в стандартном контейнере объемом 110 мл из смеси хлористого натрия и диоксида природного урана (0.294 г урана). Активность радионуклидов в препаратах рассчитывали умножением массы радиоэлемента на соответствующий коэффициент q, приведенный в последней колонке табл.1, а для определения интенсивности гамма-линий применялись квантовые выходы, приведенные в табл.3 [2,6,7].

Из табл.4 следует, что использование препаратов урана, активность которых отличается в 18 раз, дает хорошее совпадение значений эффективности (среднее отклонение менее 5% и среднеквадратичное отклонение 8%). Однако, на практике, на наш взгляд, нецелесообразно использовать для приготовления калибровочных препаратов навески урана менее 1 г, т.к. относительная погрешность, обусловленная потерями активности в процессе приготовления препарата, при малой навеске возрастает.

Для получения зависимости эффективности регистрации от энергии гамма-квантов для стандартного поглотителя (стандартной матрицы) – воды полученные значения эффективности следует разделить на коэффициент $\omega(E, \mu_m)$.

Калибровочные препараты, приготовленные смешиванием оксидов урана с неактивной матрицей, обеспечат стабильную интенсивность гамма-линий в течение длительного времени благодаря большим периодам полураспада изотопов урана. Но следует помнить, что поскольку плотность диоксида урана (2,45 г/см$^3$) существенно превышает плотность хлористого натрия (1,37 г/см$^3$), вибрации, возникающие при транспортировке и длительной эксплуатации калибровочных препаратов, могут нарушить их гомогенность, что приведет к увеличению скорости счета импульсов в регистрируемых фотопиках, хотя суммарная активность урана в препарате остается неизменной.

Т а б л и ц а 4. Результаты калибровки гамма-спектрометра с детектором GEM-30185 по эффективности

| Элемент и использованная масса, г | Радио-нуклид | Энергия гамма-линии, кэВ | Интенсивность гамма-квантов, с$^{-1}$ | Скорость счёта, имп./с | $\varepsilon$ эффективность, имп./квант | $\omega$ поправ-ка | $\varepsilon_0 = \varepsilon/\omega$ |
|---|---|---|---|---|---|---|---|
| K, 58 | $^{40}$K | 1461 | 191 | 1,608 | 0,84·10$^{-2}$ | 1,01 | 0,83·10$^{-2}$ |
| La | $^{138}$La | 788,4 | 13,8 | 0,205 | 1,49·10$^{-2}$ | 0,98 | 1,52·10$^{-2}$ |
| 66,26 | $^{138}$La | 1436 | 28,1 | 0,225 | 0,80·10$^{-2}$ | 0,99 | 0,81·10$^{-2}$ |
| U | $^{235}$U | 163,4 | 155 | 5,56 | 3,5·10$^{-2}$ | 0,87 | 4,12·10$^{-2}$ |
| 5,298 | $^{235}$U | 185,7 | 1745 | 63,79 | 3,5·10$^{-2}$ | 0,90 | 4,06·10$^{-2}$ |
| | $^{235}$U | 205,3 | 153 | 5,43 | 3,55·10$^{-2}$ | 0,92 | 3,90·10$^{-2}$ |
| | $^{234}$Th | 63,3 | 3167 | 14,52 | 0,458·10$^{-2}$ | 0,62 | 0,74·10$^{-2}$ |
| | $^{234}$Th | 92,6 | 3651 | 95,7 | 2,6·10$^{-2}$ | 0,83 | 3,2·10$^{-2}$ |
| | $^{234m}$Pa | 766,4 | 192 | 2,93 | 1,52·10$^{-2}$ | 0,98 | 1,56·10$^{-2}$ |
| | $^{234m}$Pa | 1001 | 548 | 6,26 | 1,14·10$^{-2}$ | 0,99 | 1,15·10$^{-2}$ |
| | | | | | | | |
| U | $^{235}$U | 163,4 | 8,6 | 0,296 | 3,44·10$^{-2}$ | 0,96 | 3,6·10$^{-2}$ |
| 0,294 | $^{235}$U | 185,7 | 96,84 | 3,86 | 4,0·10$^{-2}$ | 0,96 | 4,15·10$^{-2}$ |
| | $^{235}$U | 205,3 | 8,48 | 0,279 | 3,3·10$^{-2}$ | 0,97 | 3,40·10$^{-2}$ |
| | $^{234}$Th | 63,3 | 176 | 1,12 | 0,64·10$^{-2}$ | 0,79 | 0,81·10$^{-2}$ |
| | $^{234}$Th | 92,6 | 203 | 5,53 | 2,7·10$^{-2}$ | 0,91 | 3,0·10$^{-2}$ |
| | $^{234m}$Pa | 766,4 | 10,7 | 157 | 1,47·10$^{-2}$ | 0,98 | 1,50·10$^{-2}$ |
| | $^{234m}$Pa | 1001 | 30,4 | 0,316 | 1,07·10$^{-2}$ | 0,98 | 1,09·10$^{-2}$ |